\documentclass[aps,pre,twocolumn,groupedaddress,showpacs]{revtex4-1}
\usepackage{graphicx}% Include figure files
\usepackage{dcolumn}% Align table columns on decimal point
\usepackage{bm}% bold math
\usepackage{epsfig}
\usepackage{enumerate}

\usepackage{color}
\definecolor{dblue}{rgb}{0,0,0.75}
\definecolor{dred}{rgb}{0.6,0,0}
\definecolor{dgreen}{rgb}{0,0.5,0}

\input epsf
\epsfclipon

\begin{document}
\title{Oscillating hysteresis in the $q$-neighbor Ising model}
\author{Arkadiusz J\c{e}drzejewski, Anna Chmiel, Katarzyna Sznajd-Weron} 
\affiliation{Department of Theoretical Physics, Wroclaw University of Technology, Wroclaw, Poland}
\date{\today}
\begin{abstract}
We modify the kinetic Ising model with Metropolis dynamics, allowing each spin to interact only with $q$ spins randomly chosen from the whole system, which corresponds to the topology of a complete graph. We show that the model with $q \ge 3$ exhibits a phase transition between ferromagnetic and paramagnetic phases at temperature $T^*$, which linearly increases with $q$. Moreover, we show that for $q=3$ the phase transition is continuous and discontinuous for larger values of $q$. For $q>3$ the hysteresis exhibits oscillatory behavior -- expanding for even values of $q$ and shrinking for odd values of $q$. If only simulation results were taken into account, this phenomenon could be mistakenly interpreted as switching from discontinuous to continuous phase transitions or even as evidence of the so-called mixed phase transitions. Due to the mean-field like nature of the model we are able to calculate analytically not only the stationary value of the order parameter but also precisely determine the hysteresis and the effective potential showing stable, unstable and metastable steady states. The main message is that in case of non-equilibrium systems the hysteresis can behave in an odd way and computer simulations alone may mistakenly lead to incorrect conclusions. 
\end{abstract}
\maketitle

\section{Introduction}
According to the modern theory of phase transitions, each phase transition can be described by an order parameter, having a non-zero value in the ordered phase and vanishing in the disordered phase \cite{Lan:37,Fis:67,Hen:Hin:Lue:08}. The behavior of an order parameter allows to classify each phase transition as a continuous or discontinuous (first order). %From the theoretical point of view continuous phase transitions are preferably studied mainly because of their universal properties. On the other hand discontinuous transitions do occur frequently in nature \cite{Hen:Hin:Lue:08}. 
However, it has been noticed in a number of cases that the dichotomy between continuous  and discontinuous  transitions fails, in the sense that a jump of the order parameter coincides with power-law singularities \cite{Lip:00,Odo:04,Liu:Sch:Zia:12,Bar:Muk:14,She:Sha:Alv:Koe:Mac:15} or even with the absence of fundamental indicators of the first-order phase transitions, such as the hysteresis, metastable states and phase coexistence \cite{Liu:Sch:Zia:12}. It can also happen that the phase transition is weakly discontinuous, i.e. the jump of the order parameter is small and, therefore, to decide on the type of the phase transition is quite difficult in computer simulations. In such a situation measuring the hysteresis of the order parameter is a demanding task \cite{Odo:04}.  

In this paper we show that even measuring the hysteresis based solely on computer simulations can be misleading. We propose here to modify the kinetic Ising model with Metropolis dynamics \cite{God:Luc:05,Lan:Bin:13} on a complete graph by assuming that in each elementary time step a randomly chosen spin interacts only with its $q$ neighbors. We show that in such a model the hysteresis exhibits oscillatory behavior, expanding for even values of $q$ and shrinking for odd values of $q$, which could be mistakenly interpreted as switching from discontinuous to continuous phase transitions or even as evidence of the so-called mixed phase transitions \cite{Lip:00,Liu:Sch:Zia:12,Bar:Muk:14,She:Sha:Alv:Koe:Mac:15}. However, due to the mean-field like nature of the model we are able to calculate analytically not only the stationary value of an order parameter but also to determine the hysteresis and the effective potential showing stable, unstable and metastable steady states, analogously as it was done for the $q$-voter model with noise \cite{Nyc:Szn:Cis:12}.

\section{The model}
The idea to consider exactly $q$ neighbors, no matter what is the actual number of neighbors on a given graph, is borrowed from the $q$-voter model \cite{Cas:Mun:Pas:09}, originally proposed to introduce non-linearity in the voter dynamics at the microscopic level. Within the $q$-voter model, each spin is described by a dynamical variable $S_i=\pm1$ and interacts with a set of $q$ neighbors. If all $q$ neighbors share the same state, the spin conforms to this state. In the other case the spin flips with probability $\epsilon$. It is worth to notice here, that the one-dimensional $q$-voter model with $q=2$ is identical to the Ising model with generalized zero-temperature Glauber dynamics \cite{God:Luc:05}, in which a spin flips with probability $p=1$ in the case of energy decrease and with probability $p=W_0$ in the case of energy conservation. If we denote $W_0 \equiv \epsilon$, the time evolution of a single spin for both models can be written as follows:
\begin{eqnarray}
\begin{array}{l}
S_i'  = 
\left\{
\begin{array}{lll}
1 & \mbox{with} \hspace{0.2cm} p=1 & \mbox{if }  S_{i-1}=S_{i+1}=1, \\ 
-S_i & \mbox{with} \hspace{0.2cm} p=W_0 & \mbox{if } S_{i-1}S_{i+1}=-1, \\ 
-1 & \mbox{with} \hspace{0.2cm} p=1 & \mbox{if }  S_{i-1}=S_{i+1}=-1,\\
\end{array}
\right. 
\end{array}
\label{eq:in}
\end{eqnarray}
where for brevity we use the notation $S_i' \equiv S_i(t+\Delta t)$ and $S_i \equiv S_i(t)$ and $W_0=1$ corresponds to the Metropolis, whereas $W_0=1/2$ to the original Glauber dynamics \cite{God:Luc:05}. 

For higher dimensions, both models are not equivalent even in zero temperature and even for $q$ equal to the number of the nearest neighbors. The $q$-voter model requires a unanimous state of all $q$ neighbors to influence spin $S_i$, whereas for the Ising model a majority is sufficient, which follows from the Hamiltonian:
\begin{equation}
H=-\sum_{i,j}S_iS_j.
\label{eq:Ising}
\end{equation}
However, one could consider the $q$-voter model with threshold $r=1/2$ (i.e. majority needed to influence the spin) \cite{Nyc:Szn:13} and then again both models would be equivalent at zero temperature. 

The behavior of the Ising model described by the Hamiltonian in Eq. (\ref{eq:Ising}) under zero-temparature Glauber dynamics is very interesting, exhibiting a slow relaxation related to a metastable state \cite{God:Luc:05,Spi:Kra:Red:01,Ole:Kra:Red:11}. However, here we focus on another problem related to the kinetic Ising model, inspired by the analogy between the one-dimensional kinetic Ising model with zero-temperature Glauber dynamics and the $q$-voter model with $q=2$. We ask the following question: \emph{what would be the behavior of a modified kinetic Ising model -- in which every spin interacts with a set of $q$ neighbors randomly chosen from the set of all its neighbors -- if we introduced a temperature-like parameter $T>0$?} 

The algorithm of a single time step of the $q$-neighbor Ising model consists of 3 consecutive steps: 
\begin{enumerate}
\item Randomly choose a spin, $S_i$, and from all its neighbors choose a subset of $q$ neighbors, $nn_q$. 
\item Calculate the value of the following function, based on the Hamiltonian in Eq. (\ref{eq:Ising}), for the original state of the $i$-th spin:
\begin{equation}
E(S_i)=-S_i\sum_{j \in nn_q}S_j,
\label{eq:qIsing}
\end{equation}
and the value of the same function for the flipped $i$-th spin, i.e. $E(-S_i)$.
\item Flip the $i$-th spin with probability $\min[1,e^{-\Delta E/T}]$, where $\Delta E=E(-S_i)-E(S_i)$. 
\end{enumerate}
We would like to stress here, that minimizing the function given in Eq. (\ref{eq:qIsing}) does not necessarily lead to the minimization of the whole energy of the system given by Eq. (\ref{eq:Ising}), in contrast to the equilibrium Ising model in which we sum interactions over all nearest neighbors. 
%Here we calculate $E(S_i)$ taking into account only $q$ neighbors. This kind of function
%A function like $E(S_i)$, which controls the flip of the spin being minimized only locally, has been called the disagreement function \cite{Szn:04}. 
As usual we choose the magnetization:
\begin{equation}
m(t)=\frac{1}{N}\sum_{i=1}^N S_i(t),
\label{eq:mag}
\end{equation}
as an order parameter, which in the case of a complete graph fully describes the state of the system.

\section{Results}

We investigate the model using an analytical approach and Monte Carlo simulations. The latter start from two types of initial conditions -- fully ordered ($m=1$), which corresponds to zero temperature, and completely random ($m=0$), which corresponds to a high temperature. For each value of the temperature-like parameter $T$, we measure the stationary value of the magnetization defined by Eq. (\ref{eq:mag}). We have checked that averaging over time gives the same result as averaging over samples. However, when using the time average it is easier to distinguish between continuous and discontinuous phase transitions looking solely at the order parameter as a function of $T$. Starting from two types of initial conditions should allow us to identify the type of the transition on the basis of the hysteresis, but as we will see later, the hysteresis can be also misleading. 

\begin{figure}
%\vskip 0.3cm
 \centerline{\epsfig{file=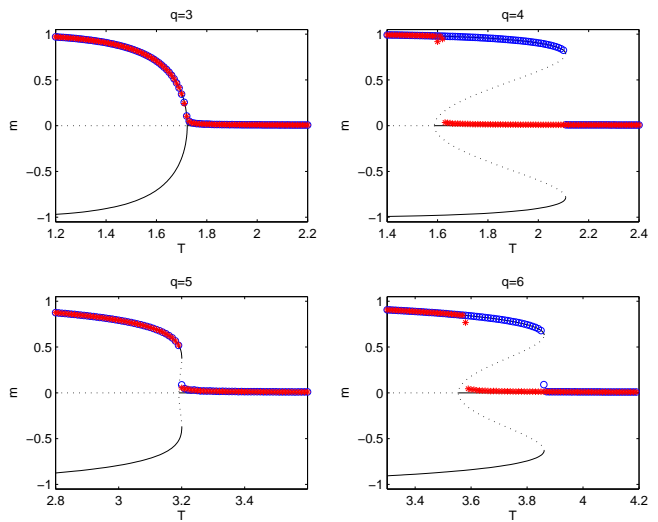,width=1.0\columnwidth}}
  \caption{Dependencies between steady values of the magnetization $m$ and the temperature-like parameter $T$ for 4 values of $q$. Symbols represent results obtained from Monte Carlo simulations for a system of size $N=10^5$ and two types of initial conditions -- fully ordered (o) and disordered (*). Numerical results, see Eq. (\ref{F0}), are presented by lines -- solid for stable solutions and dotted for unstable.}
\label{fig_histeresis}  
\end{figure}

Dependencies between steady values of the magnetization $m$ and the temperature-like parameter $T$ are presented in Fig. \ref{fig_histeresis}. A phase transition between ordered and disordered phases is observed for all values of $q \ge 3$. For $q=3$ there is no jump in the order parameter and no hysteresis, which indicates a continuous phase transition. For $q=4$ and $q=6$ the jump of the order parameter and the hysteresis indicate a discontinuous phase transition. However, in the case of $q=5$, we are not able to distinguish between the continuous and discontinuous phase transition -- the jump of the order parameter is observed and simultaneously there is no hysteresis, similarly as observed in \cite{Liu:Sch:Zia:12}. 

Fortunately, thanks to the mean-field type of the model we are able, following the reasoning presented in \cite{Szn:04,Nyc:Szn:Cis:12}, to write down equations that allow to calculate the stationary value of the order parameter, as well as an effective potential. The latter helps to distinguish between continuous and discontinuous phase transitions.

Although the magnetization $m$ is an order parameter of the system and all results will be presented in terms of $m$, calculations are easier if we use the concentration of `up-spins' which is equivalent to the probability that a randomly chosen spin is `up'. In a single time step $\Delta_t$, three events are possible -- the concentration of `up-spins' $c$ increases by $1/N$, decreases by $1/N$ or remains constant:
\begin{eqnarray}
\gamma^+(c,T,q) & = & Prob\left\{c \rightarrow c+\Delta_N \right\}, \nonumber\\
\gamma^-(c,T,q) & = & Prob\left\{c \rightarrow c-\Delta_N \right\}, \nonumber\\
\gamma^0(c,T,q) & = & Prob\left\{c \rightarrow c\right\} = 1-\gamma^+(c)-\gamma^-(c),
\end{eqnarray}
where: 
\begin{eqnarray}
\gamma^+(c,T,q) = \sum_{k=0}^{q}{{q \choose k}c^{q-k}(1-c)^{k+1}} \min \left[ 1, e^{\frac{2}{T}(q-2k)} \right], \nonumber \\
\gamma^-(c,T,q) =  \sum_{k=0}^{q}{{q \choose k}(1-c)^{q-k}c^{k+1}} \min \left[ 1, e^{\frac{2}{T}(q-2k)} \right]. \nonumber \\
\end{eqnarray}

In the stationary state we expect that the probability of growth $\gamma^+(c,T,q)$ is equal to the probability of loss $\gamma^-(c,T,q)$. Therefore:
\begin{equation}\label{F0}
F(c,T,q)=\gamma^+(c,T,q)-\gamma^-(c,T,q)=0, 
\end{equation}
where $F(c,T,q)$ can be treated as an effective force that drives the concentration $c$ up or down \cite{Nyc:Szn:Cis:12}. 
%To calculate stationary values of concentration we have to solve the equation:
%\begin{equation}
%F(c,T,q)=0.
%\end{equation}
Solving analytically Eq. (\ref{F0}), i.e. finding the stationary value of $c$ as a function of $T$ for an arbitrary value of $q$, is impossible, but we can easily do it numerically. The results are denoted by lines in Fig. \ref{fig_histeresis}. 

\begin{figure}
%\vskip 0.3cm
 \centerline{\epsfig{file=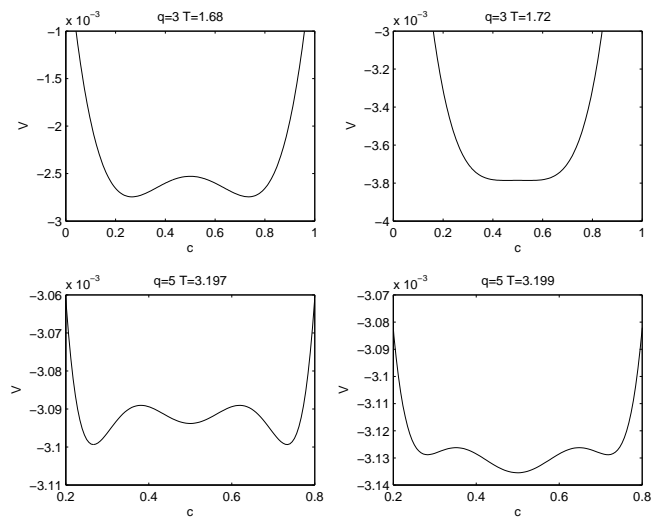,width=1.0\columnwidth}}
  \caption{Potentials given by Eq. (\ref{pot}) for $q=3$ (top panels) and $q=5$ (bottom panels). Left panels represent potentials just below and right panels just above the transition point. It is seen that for $q=3$ there are two ordered phases below the transition point and a single disordered phase above the transition point, which is typical for continuous phase transitions. For $q=5$ the phase coexistence and metastable states are seen, which indicates a discontinuous phase transition.}
\label{fig_pot_q3q5}  
\end{figure}

Having the effective force $F(c,T,q)$ we can also calculate the effective potential:
\begin{eqnarray}
V(c,T,q) = - \int F(c,T,q) dc, 
\label{pot} 
\end{eqnarray}
which, as seen in Fig. \ref{fig_pot_q3q5}, allows to distinguish between stable (minima of the potential) and unstable (maxima of the potential) solutions of Eq. (\ref{F0}).  Furthermore, it allows to distinguish between continuous and discontinuous phase transitions, in the latter case showing phase coexistence and metastable states. Finally, it allows to determine the transition point which in case of a discontinuous phase transition coincides with the value of $T$, for which minima corresponding to disordered and ordered phases are equal. 

\begin{figure}
%\vskip 0.3cm
 \centerline{\epsfig{file=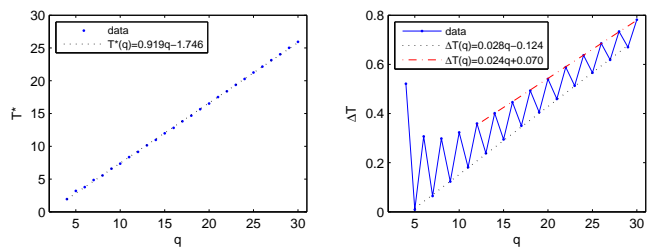,width=1.0\columnwidth}}
  \caption{Dependence between the phase transition value of $T$ (i.e. $T^*$; \emph{left panel}) or the width of the hysteresis (i.e. $\Delta T$; \emph{right panel}) and the number of neighbors (i.e. $q$).}
\label{fig_hist}  
\end{figure}

Having the effective force (\ref{F0}) and the effective potential (\ref{pot}) we can calculate numerically not only the stationary value of the order parameter $m(T)$ for an arbitrary value of $q$, but also determine the transition temperature $T^*$ and the width of the hysteresis defined here as the distance between the spinodal lines (see Fig. \ref{fig_hist}). 

%Already looking at the left panel of Fig. \ref{fig3} we see that hysteresis for odd values of $q$ is narrower than for even values of $q$. Moreover, it is seen in the right panel of Fig. \ref{fig3}, hysteresis widens with $q$. 
%
%\begin{figure}
%\vskip 0.3cm
 %\centerline{\epsfig{file=2X1panel_HistEven20_30.eps,width=1.0\columnwidth}}
  %\caption{Dependencies between steady values of the magnetization $m$ and the temperature-like parameter $T$ for different values of $q$ obtained from the Eq. (\ref{F0}) are presented by lines (solid lines for stable solutions and dashed for unstable).}
%\label{fig_3}  
%\end{figure}

More precisely, the latter can be calculated from the potential (\ref{pot}). For low values of $T$ there are two minima that correspond to ordered phases. Then at $T=T_1$ the third minimum appears; it corresponds to the disordered phase but it shallower than other two, i.e. the disordered state is metastable. At $T=T^*$ all three minima are equal, which corresponds to the transition point. Above this value the middle minimum corresponding to the disordered phase is the deepest and the other two represent metastable ordered states. Finally, above $T=T_2$ there is only one minimum -- the disordered state is the only possible state of the system. The distance between the spinodal lines $\Delta T=T_2-T_1$ determines the width of the hysteresis and is presented in the right panel of Fig. \ref{fig_hist}.

In Fig. \ref{fig_hist} it can be seen that the transition temperature $T^*$ increases linearly with $q$, whereas the hysteresis exhibits an oscillatory behavior, expanding for even values of $q$ and shrinking for odd values of $q$. Interestingly, for $q=5$ the width of the hysteresis is roughly zero and even for $q=7$ it is difficult to see any hysteresis in computer simulations. This fact has initially led us to the wrong conclusion of oscillatory switching from continuous to discontinuous phase transitions.

\section{Conclusions}

It is believed that the hysteresis is the main indicator of discontinuous phase transitions \cite{Hen:Hin:Lue:08,Odo:04}. However, as we have shown here, the behavior of the hysteresis might be quite unexpected. %It may mistakenly lead to an incorrect conclusion of a hybrid phase transition, which has been recently observed in a number of systems \cite{Lip:00,Liu:Sch:Zia:12,Bar:Muk:14}. 
In the simple model studied here, we have observed a jump of the order parameter and simultaneously no hysteresis for $q=5$.  Even for larger odd values of $q$ ($q=7,9$) the hysteresis was invisible in computer simulations. If the model was not solvable analytically, the results of computer simulations could drive us to the wrong conclusion of a hybrid phase transition in which the jump of the ordered parameter coincides with no hysteresis. Luckily, due to the mean-field character of the model, we could distinguish between the two types of transitions on the basis of the effective potential. %In the case of discontinuous phase transitions this clearly shows a phase coexistence and metastabilities.  

In equilibrium statistical mechanics, it is common that systems, which exhibit a discontinuous phase transition in high space dimensions, may display a continuous transition below a certain upper critical dimension \cite{Hen:Hin:Lue:08}. We expect that at the same time the hysteresis monotonically decays, reaching zero at the upper critical dimension. Here we have similar situation, despite the fact that we do not change the dimension but only the number of neighbors $q$. For $q>3$ the transition is discontinuous, for $q=3$ continuous and for $q<3$ there is no phase transition in the system. However, at the same time the hysteresis does not decay monotonically but oscillates. Therefore, the main message of our paper is a warning that in case of non-equilibrium systems the hysteresis can behave in an odd way and computer simulations alone may mistakenly lead to incorrect conclusions. 

\begin{acknowledgments}
This work was supported by funds from the National Science Centre (NCN, Poland) through post-doctoral fellowship no. 2014/12/S/ST3/00326 (to AC) and grant no. 2013/11/B/HS4/01061 (to AJ and KSW).
\end{acknowledgments}

\end{document}